\documentclass[twocolumn,trackchanges,apjl,resetfootnote]{aastex701}
\usepackage{amsmath}
\usepackage{booktabs}
\usepackage{graphicx}
\begin{document}

\title{Atmospheric collapse and re-inflation through impacts for terrestrial planets around M dwarfs}

\author[orcid=0000-0003-3829-8554, gname=Prune Camille,sname=August]{Prune C. August}
\affiliation{Technical University of Denmark, Department of Space Research and Technology}
\affiliation{Harvard University, Department of Earth and Planetary Sciences}
\email{prua@space.dtu.dk}

\author[0000-0003-1127-8334]{Robin Wordsworth}
\affiliation{Harvard University, School of Engineering and Applied Sciences}
\affiliation{Harvard University, Department of Earth and Planetary Sciences}
\email{rwordsworth@seas.harvard.edu}

\author[0009-0000-2828-2263]{Mikayla Huffman}
\affiliation{University of Colorado Boulder, Astrophysical and Planetary Sciences Department}
\email{mikayla.huffman@colorado.edu}

\author[0000-0001-8932-368X]{David Brain}
\affiliation{University of Colorado Boulder, Astrophysical and Planetary Sciences Department}
\email{david.brain@colorado.edu}

\author[0000-0003-1605-5666]{Lars A. Buchhave}
\affiliation{Technical University of Denmark, Department of Space Research and Technology}
\email{buchhave@space.dtu.dk}

\begin{abstract}
Detection of an atmosphere around a terrestrial exoplanet will be a major milestone in the field, but our observational capacities are biased towards tidally locked, close-in planets orbiting M-dwarf stars. The atmospheres of these planets are vulnerable to atmospheric erosion and collapse due to condensation of volatiles on the nightside. However, these condensed volatiles constitute a stable reservoir that could be re-vaporised by meteorite impacts and re-establish the atmospheres. Through a simple energy balance model applied to atmospheric evolution simulations with stochastic impacts, we assess the viability and importance of this mechanism for CO$_2$ atmospheres. We find that moderate-sized impactors ($5-10~\rm{km}$ diameter) occurring at a frequency of $1-100~\rm{Gyr}^{-1}$ can regenerate observable transient atmospheres on previously airless planets. We focus on specific targets from the JWST DDT Rocky Worlds programme, and compute the fraction of their evolution spent with a transient CO$_2$ atmosphere generated through this mechanism. We find this fraction can reach $70\%$ for GJ~3929~b, $50\%$ for LTT~1445~Ac, $80\%$ for LTT~1445~Ab, at high impact rates and strong CO$_2$ outgassing over the planet's lifetime. We also show that atmospheric collapse can shield volatiles from escape, particularly in the early, high-XUV phase of M-dwarf evolution. Overall, our work suggests that terrestrial planet atmospheres may not evolve monotonically but instead may be shaped by episodic external forcings.

\end{abstract}

\section{Introduction}
The influence of meteorite impacts on atmospheric evolution of terrestrial planets is well documented. Incoming comets can deliver volatiles to replenish atmospheres \citep[e.g.][]{Morbidelli2000, Ciesla2015} or conversely, erode them through high energy collisions \citep[e.g.][]{Shuvalov2009, Kral2018}. They can also dramatically alter the climatic conditions of a planet by vaporising target materials and injecting them into the atmosphere, as exemplified by the Chicxulub impact \citep{Alvarez1980, Pierazzo1998}. On Mars, impact cratering has been proposed as a mechanism to trigger transient warming and precipitation necessary to explain the observed extensive valley networks \citep{Segura2002, Segura2008, Toon2010}. Building on this idea, \cite{PalumboHead2018} further quantified the water cycle that would result from such an impact, showing that large and basin-scale impacts would result in high and homogeneously distributed rainfall. Although the authors conclude the mechanism is incompatible with the observed equatorial concentration of valley networks, their work demonstrates the capacity of impacts to regenerate significant transient atmospheres. 

Collapsed or partially collapsed atmospheres are common in the Solar System. Mars' polar ice caps likely have been formed and shaped by collapse processes \citep[][]{Forget2013, Soto2015}, and MESSENGER observations revealed evidence of volatile-rich surfaces on Mercury, where evidence of volatile-rich layers and glacier-like terrains has also been found \citep[e.g.][]{Rodriguez2023}. Additionally, spectra of Io's atmosphere before and during its eclipse by Jupiter show the SO$_2$ feature appearing and disappearing, indicating that the atmosphere collapses on its surface when in Jupiter's shadow and gets reinflated as it comes back out \citep{Tsang2016}. These examples illustrate that collapsed atmospheres can act as volatile reservoirs, capable of regenerating transient atmospheres if sufficient energy is provided.

Cratering records on the Moon  \citep{Shoemaker1969,Shoemaker1970}, Mercury \citep{Strom2008} and Mars \citep{HartmannNeukum2001} suggest that impacts are a common phenomenon in the Solar System, at least on geological timescales. On Earth, estimated impact rates are 10 to 100 impacts/Gyr for objects with a diameter $>5~\rm{km}$ and 1 to 10 impacts/Gyr for those $>10~\rm{km}$ \citep{Binzel2004, LeFeuvre2011}. Impactor fluxes are currently poorly constrained on exoplanets, but TESS photometric observations of exocomet transits in the $\beta$ Pic b system indicate a differential power law slope of $\gamma = 3.6 \pm 0.8$ for the comet size distribution, consistent with both Solar System comet populations and with expectations for a collisionally relaxed distribution ($\gamma = 3.5$) \citep{Lecavelier2022}. 

Close-in tidally locked rocky exoplanets around M dwarfs may host substantial surface ice reservoirs in cold traps on their nightsides. These planets experience a high amount of XUV radiation, which significantly erodes their atmospheres \citep[e.g.,][]{Shields2016, Krissansen2023,WordsworthKreidberg2022}. Once the atmospheric pressure drops below a critical threshold, the nightside becomes cool enough for the volatiles to condense on the surface, initiating catastrophic atmospheric collapse \citep{Wordsworth2015}. Once the atmosphere has collapsed, a new stable state is reached, and continued outgassing leads to further condensation on the nightside. \footnote{The only exception would be if the mass outgassed exceeded that of the critical pressure threshold of the planet over timescales shorter than condensation timescales, which is unlikely from volcanism alone.} This creates the possibility of vast volatile reservoirs trapped in ice sheets on the nightside, which could in principle be vaporised by impact heating \citep{PalumboHead2018,Turbet2020}.

The launch of the James Webb Space Telescope (JWST) has given a major push to the hunt for terrestrial exoplanet atmospheres. Many programmes targeting M-dwarf rocky planets have returned non-detections \citep[e.g.,][Allen et al. submitted]{Zieba2023, Zhang2024, Xue2024, Mansfield2024, MeierValdes2025, Fortune2025}, or at most tentative detections \citep[][]{Hu2024, August2025, BelloArufe2025}, with many of these results requiring additional observations to establish a consensus. The recently announced 500 hour Director's Discretionary Time (DDT) "Rocky Worlds" programme \citep{Redfield2024} underscores the timeliness of theoretical work on the formation and survival of secondary atmospheres on these planets.

Here, we propose a mechanism for the formation of transient, secondary atmospheres around rocky exoplanets. Section \ref{sec:theory} outlines the theoretical basis for atmospheric collapse and our impact-driven regeneration mechanism, and Section \ref{sec:model} describes our model. In Section \ref{sec:results}, we show the results of our simulations, and we discuss key findings and limitations in Section \ref{sec:discussion}. Finally, we summarise and link our findings to the current observational context in Section \ref{sec:conclusion}, and propose avenues for future work.

\section{Theory}\label{sec:theory}

\begin{figure*}
    \centering
    \includegraphics[width=0.6\linewidth]{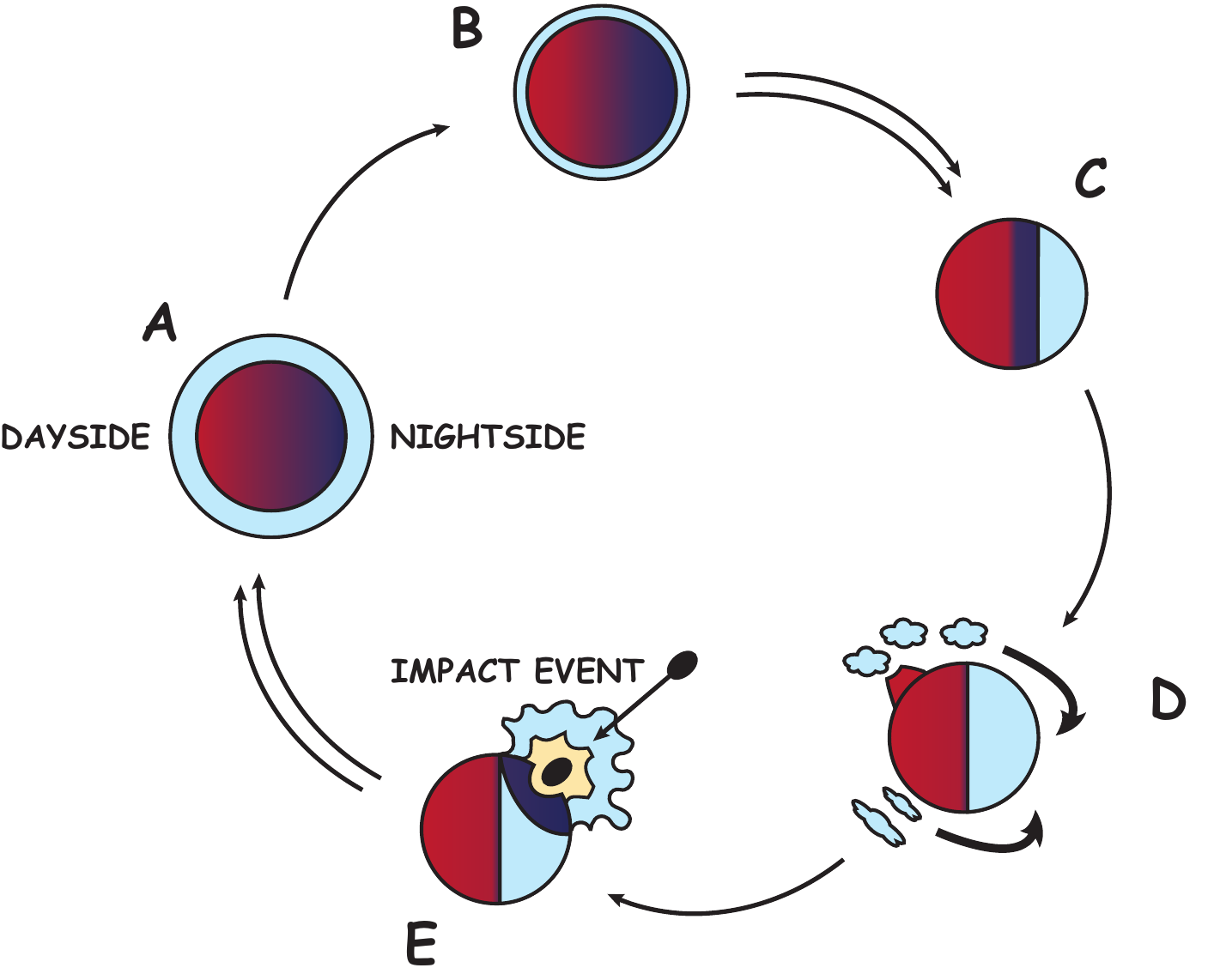}
    \caption{Schematic of episodic atmospheric collapse and regeneration through impacts for a tidally locked planet. A) The planet has a volatile rich atmosphere which redistributes heat from the day- to the nightside. B) Atmospheric escape thins the atmosphere, heat redistribution becomes less efficient, nightside temperatures drop. C) Nightside temperatures have reached the volatile condensation temperature, the atmosphere collapses. D) Volatiles outgassed through volcanism or magma ocean pockets accumulate on the nightside as ice. E) An impactor hits the nightside and vaporises ice and rock. Hot vapour, ejecta, and silicate rain further vaporise the nightside ice sheets (see Section~\ref{subsec:th_reinf} for details). An atmosphere is regenerated.}
    \label{fig:schematic}
\end{figure*}

\subsection{Atmospheric collapse}\label{subsec:th_atmcoll}
Atmospheric collapse on tidally locked exoplanets was first proposed in the 1990s \citep{Kasting1993,Joshi1997}, with the physics of the process first studied in detail starting in the 2010s \citep{Wordsworth2015,Koll2016}. 
Atmospheric collapse occurs when the atmosphere is too thin to warm the planet's coldest regions sufficiently to prevent condensation of the main constituent on the surface. Once collapse begins, a positive feedback cycle resulting in complete collapse onto the surface has occurred (steps B to C in Figure~\ref{fig:schematic}). The critical pressure at which this happens is given by the balance between the temperature of the surface and the condensation temperature of the volatiles making up the atmosphere. 

Most terrestrial planets suitable for observations orbit M dwarfs at short orbital separations, which makes them likely to experience atmospheric collapse. Firstly, this proximity to their host star drives significant atmospheric erosion through exposure to high XUV fluxes \citep[e.g.,][]{Tian2009, Medina2020}, flares \citep[][]{Medina2022, doAmaral2022}, and stellar winds \citep{Dong2018}. The light, primary H/He atmospheres are likely to be fully eroded, leaving heavier molecules like CO$_2$ as strong candidates for secondary atmospheres \citep[e.g.,][]{Krissansen2023}. Secondly, close-in planets are likely to be tidally locked \citep[e.g.,][]{Kasting1993}, favouring condensation of volatiles on the nightside. For planets that are not tidally locked, this collapse can also occur at the poles if obliquity is low.

In this work, we follow the approach of \cite{Wordsworth2015} and focus on CO$_2$ atmospheres on tidally locked planets. CO$_2$ is not only a plausible main constituent of hot rocky exoplanet atmospheres \citep[e.g.,][]{Tian2009} but also spectroscopically active in the infrared, making it observationally relevant for atmospheric characterisation missions with JWST (e.g., the \textit{Hot Rocks Survey}, \citealt{DiamondLowe2023prop}; \textit{Rocky Worlds}, \citealt{Redfield2024}). The condensation temperature of CO$_2$ as a function of pressure is given by: 
\begin{equation}\label{eq:tcondco2low}
    T_{\rm{cond,CO}_2}(p) = \frac{-3167.8}{\ln{0.01p} - 23.23},
\end{equation}
for $p<51800~\rm{Pa}$, which corresponds to the triple pressure point, and 
\begin{equation}\label{eq:tcondco2high}
    T_{\rm{cond,CO}_2}(p) = 684.2 - 92.3\ln{p} + 42.3\ln{p}^2,
\end{equation}
for $p>51800~\rm{Pa}$ \citep{Fanale1982, Wordsworth2010}.
The surface temperature of the nightside is computed using the thin radiator approximation from \citet{Wordsworth2015} :
\begin{equation}
    T_n(p) \approx T_{\rm{tr}}(p) \equiv \Bigg(\frac{(1-A)F \kappa p}{4\sigma g}\Bigg)^{1/4},
\end{equation}
where $A$ is the surface albedo, $F$ is the incoming stellar flux, $g$ is the gravity, $\kappa$ is the infrared atmospheric opacity, and $\sigma$ is the Boltzmann constant. In our work, we use a surface albedo of $A=0.2$, and $\kappa = 1.6\cdot 10^{-4}~\rm{m^2~kg^{-1}}$, which was specifically picked to approximately match the nightside temperatures produced by the simulations for CO$_2$ atmospheres at $0.1~\rm{bar}$.

\subsection{Reinflation of a collapsed atmosphere}\label{subsec:th_reinf}
The process through which large impacts can regenerate atmospheres has many steps \citep{Segura2008}. The main contribution comes from the vaporisation of surface and subsurface ice \citep{Melosh1989}. The impact itself causes part of target material to vaporise, creating a vapour plume of both volatile and rock \citep{SleepZahnle1998}. As the plume cools, it radiates heat towards the surface, causing more melting and sublimation of surface and subsurface ice. When the plume cools sufficiently, the silicates condense and rain out, forming a global silicate condensate layer that further heats surface volatiles over a large area \citep{Segura2002}. Finally, a hydrologic cycle may result as water from the plume condenses, forming rain, which then re-evaporates upon contact with the hot silicate condensate layer \citep{PalumboHead2018}. In the case of early Mars, this cycle could have continued until the layer cooled enough and the atmosphere returned to an ambient state.

The timescale of the processes contributing to reinflation are on the order of a few hundred years, meaning they can be considered instantaneous for our purposes (step E of the schematic in Figure~\ref{fig:schematic}). We can also assume that once a mass of volatile ice equivalent to the critical pressure (defined in Section~\ref{subsec:th_atmcoll}) has been vaporised, the resulting atmospheric warming triggers vaporisation of the remaining surface volatile ice, in a reverse of the positive feedback described in Section~\ref{subsec:th_atmcoll}. 

Not all impacts trigger reinflation. As well as striking the planet's nightside where the volatile ices are located, the impactor must supply sufficient total energy to vaporise a critical mass of volatile ice.  
The kinetic energy of an incoming impactor is
\begin{equation}\label{eq:ekin}
    E_{kin}^{imp} = \frac{1}{2}m_{imp}v_{imp}^2.
\end{equation}
where $m_{imp}$ and $v_{imp}$ are the mass and velocity of the impactor. 
The speed of the impactor on impact depends on the planet's gravitational potential as well as the impactor's excess speed on entering the planet's sphere of influence.  The effective minimum speed of an impactor hitting an airless planet is the escape velocity of the planet, $v_{esc}$. Here we define $v_{imp} \equiv v_{esc}$, using the escape velocity as a conservative lower bound on the incoming impactor's speed.

The energy necessary to vaporise a critical mass of volatiles, $m_{c} = {4\pi R_p^2 p_{c}}/{g}$, is given by :
\begin{equation}
    E_{vap} = m_{c}l_{\rm{CO_2}} + m_{c}c_{p,\rm{CO_2}}\Delta T
\end{equation}
where $l_{\rm{CO_2}}$ is the latent heat, $c_{p,\rm{CO_2}}$ the specific heat capacity and $\Delta T = T_{sub, \rm{CO_2}} - T_{surf}$ is the difference between the sublimation temperature and the surface temperature. The nightside temperature is likely to be extremely low, so we conservatively set $\Delta T = T_{sub, \rm{CO_2}}$. For reference, permanently shadowed craters on Mercury and the Moon reach only 40-50K \citep{Mukai1997}. Additionally, the choice of $T_{surf}$ has a minor impact on the outcome since the specific heat term is about an order of magnitude smaller than the latent heat term for CO$_2$.

Assessing how much of the energy from the impact is actively converted into energy to vaporise the target material is a potentially complex problem. Here, for simplicity, we use an efficiency parameter $\epsilon$ to represent the transfer of kinetic energy to thermal energy via shock physics. \cite{AhrensOKeefe1972, AhrensOKeefe1977} studied the energy partitioning of a meteorite impact on iron and gabbroic anorthosite surfaces and find values ranging between $\epsilon = 0.7$ to $0.9$ depending on the impact velocity as well as the meteorite and target surface material. For impacts on H$_2$O ice, using the scaling laws from \citet{Kraus2011}, we find shock vaporisation efficiencies ranging from $\epsilon = 0.1-0.3$, depending on impact velocity and target porosity. We take a representative value of $\epsilon = 0.5$ in our model. In reality, several competing factors may influence $\epsilon$. Geometric effects from three-dimensional heat redistribution during cratering \citep[e.g.][]{Barnhart2011} could reduce vaporisation efficiency, although rock vapour, hot ejecta and volatiles transport heat over a much larger area \citep{Segura2002,Toon2010}. Quantifying which mechanism dominates is beyond the scope of this work, and does not significantly affect our conclusions unless the process became highly inefficient ($\epsilon \ll 0.01$).

\begin{figure}
    \centering
    \includegraphics[width=\linewidth]{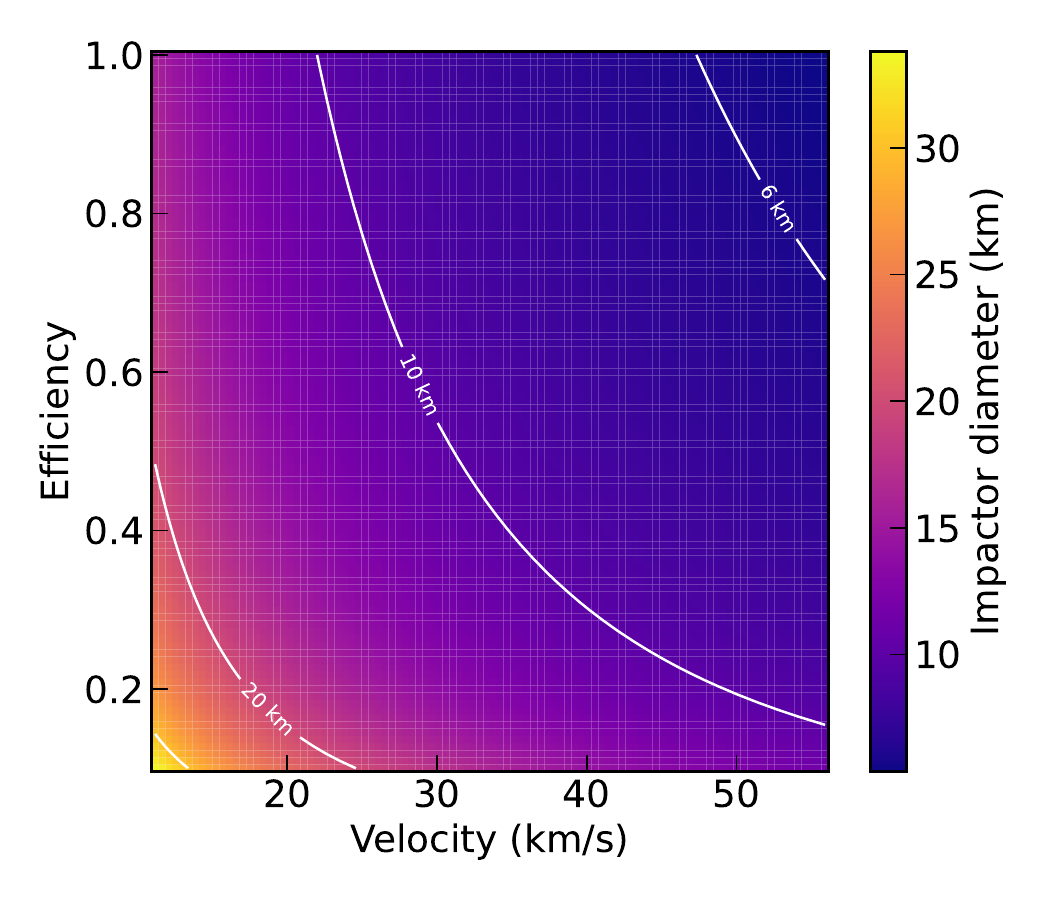}
    \caption{Minimum diameter of an impactor necessary to trigger reinflation on an Earth-like planet for different values of $\epsilon$ and $v_{imp}$, assuming $\rho_{imp} = 3.0~\rm{g~cm^{-1}}$.}
    \label{fig:mindiameter}
\end{figure}

Combining to get a constraint on the minimum mass of an impactor $m_{imp,min}$ for a given planet with an escape velocity $v_{esc}$, we need
\begin{equation}
    \epsilon E_{kin}^{imp} \geq E_{vap}
\end{equation}
and hence
\begin{equation}\label{eq:massimpactor}
    m_{imp,min} \equiv \frac{2E_{vap}}{\epsilon v_{esc}^2}.
\end{equation}
In this work, we express the impactor dimensions using an equivalent diameter. The equivalent diameter is the diameter the impactor would have for a given mass if it were perfectly spherical, that is :
\begin{equation}\label{eq:diameter}
    D_{imp} = 2\Bigg(\frac{m_{imp}}{\frac{4}{3}\pi\rho_{imp}}\Bigg)^{1/3}
\end{equation}
For example, in order to reinflate a $0.1~\rm{bar}$ atmosphere of CO$_2$ on Earth, assuming an impactor density of $\rho_{imp} = 3.0~\rm{g~cm^{-1}}$, the minimum diameter is about $20~\rm{km}$. However, this can drop to $6~\rm{km}$ for higher velocities and efficiencies (see Figure~\ref{fig:mindiameter}).

Choices of particular values for $\epsilon$, $v_{imp}$, $\rho_{imp}$ are interchangeable within this energy budget framework. Due to the square dependency on velocity in the kinetic energy, an increase in impact velocity makes for proportionally smaller impactor masses and diameters to trigger reinflation. The efficiency of vaporisation has a direct linear relationship with minimum impactor mass. We therefore fix both variables to reasonable, but conservative, values for order-of-magnitude estimates of this physical process. Figure \ref{fig:mindiameter} shows this dependency for Earth's reinflation at $0.1~\rm{bar}$. The minimum diameter drops from $20~\rm{km}$ ($v_{imp} = v_{esc,\oplus}$ and $\epsilon = 0.5$) to $6~\rm{km}$ at higher impact velocities and efficiencies.

\section{Model}\label{sec:model}
\subsection{Evolution of a two-state system}\label{subsec:2state}
We model a single planet's atmospheric evolution as a two-state system\footnote{Code available at \url{https://github.com/prunecamille/impact-reinflation-code} and archived on Zenodo \citep{August2025code}.}. In the inflated state, the atmospheric reservoir experiences loss, driven by the host star's XUV irradiation, as well as replenishment, through outgassing. In the collapsed state, atmospheric escape effectively stops, and outgassed volatiles immediately condense on the nightside. This behaviour is captured by the following set of equations :
\begin{equation}\label{eq:evolution}
    \begin{cases}
       \frac{dM_{\rm{atm}}}{dt} = \phi_{\rm{out}}(t) - \phi_{\rm{esc}}(t), & p \geq p_c; \\
       \frac{dM_{\rm{atm}}}{dt} = \phi_{\rm{out}}(t), & p < p_c.
    \end{cases}
\end{equation}
where $\phi_{\rm{esc}}$ and $\phi_{\rm{out}}$ are the CO$_2$ escape and outgassing fluxes respectively. We solve Equation~\eqref{eq:evolution} via the Euler method. Runs with a timestep size $\delta t = 10^6~\rm{yr}$ and $\delta t = 10^5~\rm{yr}$ produced consistent results, so we adopt $\delta t = 10^6~\rm{yr}$ for efficiency and evolve the planets over several gigayears.

The atmospheric escape term is modelled using the formula for energy-limited escape \citep{Watson1981} :
\begin{equation}\label{eq:escape}
    \phi_{\rm{esc}} = \frac{\eta F_{XUV}(t)}{4V_{pot}}
\end{equation}
where $F_{XUV}(t)$ is the time-dependent XUV-evolution of the star and $V_{pot}$ the gravitational potential of the planet.
We set the efficiency to $\eta=0.01$, informed by CO$_2$ escape rate estimates from \cite{Tian2009}; Section~\ref{subsec:disc_escout} discusses our rationale for this choice.

For the XUV-evolution of the star, we follow `model A' outlined in \citet{Wordsworth2018}. The saturation phase luminosity is set to $F_{sat} = 10^{-3} F_{\rm{bol}}$, and the time evolution after the saturation phase is modelled as a decaying power law \citep{Ribas2005}:
\begin{equation}\label{eq:fxuv}
    \begin{cases}
        F_{XUV}(t) = F_{sat}, & \text{for } t < t_{sat} \\
        F_{XUV}(t) = F_{sat}\big(\frac{t}{t_{sat}}\big)^{-1.23}, & \text{for } t \geq t_{sat}
    \end{cases}
\end{equation}
The duration of the saturation phase, $t_{sat}$, is set individually for each system, for a more accurate representation of M-dwarf evolution across spectral types.

The outgassing term is scaled as constant factor of present-day Earth CO$_2$ outgassing rates. \citet{WallmannAloisi2012} report values on the order of $5$ to $10\times 10^{12}~\rm{mol~yr^{-1}~C}$, which we convert to a mass rate per unit area of $4.3$ to $8.6\times 10^{-4}~\rm{kg~m^{-2}~yr^{-1}}$ of CO$_2$. This allows the outgassing rate to be scaled according to the surface area of each planet. To account for uncertainties in CO$_2$ outgassing rates across planetary bodies, we adopt a representative value for Earth's present day outgassing of $6.5\times 10^{-4}~\rm{kg~m^{-2}~yr^{-1}}$ as our reference and explore a range spanning 0.01 to 2 times this value. This range is chosen to span plausible variations while acknowledging that our simulations neglect carbon cycling processes \citep[such as silicate weathering and carbonate formation,][]{Walker1981} that regulate atmospheric CO$_2$ on Earth.
More discussion on the outgassing term can be found in Section~\ref{subsec:disc_escout}.

The system remains inflated for as long as the atmospheric pressure exceeds $p_{c}$; it then falls into the collapsed state. The system can only switch back into an inflated state through the action of a sufficiently energetic impact.

\subsection{Stochastic impacts}\label{subsec:stoch}
Impacts are modelled as stochastic events following a Poisson distribution with an impact rate $\tau$. With a fixed impact velocity and efficiency as described in Section \ref{subsec:th_reinf}, any impactor larger than the selected impactor diameter (see Equation~\ref{eq:diameter}) will trigger a reinflation.

In the absence of detailed modelling of impactor populations for specific exoplanets, we explore the parameter space by picking the minimum impactor size necessary to trigger reinflation for a specific planet, and use the impact rate as our control variable. This approach is effectively equivalent to asking how often do impact events with energy exceeding $E_{vap}/\epsilon$ occur over an exoplanet's lifetime. We further discuss impact rate in Section~\ref{subsec:disc_impactrates}. In every run, a successful impact instantaneously returns the system to an inflated state by re-injecting all the volatiles accumulated on the nightside into the atmosphere. 

We stress that the impact events are stochastic, meaning each run of the simulation produces different evolutions for the same set of initial planetary and stellar conditions. This also means that a fixed impact rate does not necessary result in exactly the corresponding amount of impacts over the simulation for a single run, but only on average over a large number of runs.

\begin{figure*}
    \centering    \includegraphics[width=0.8\linewidth,clip]{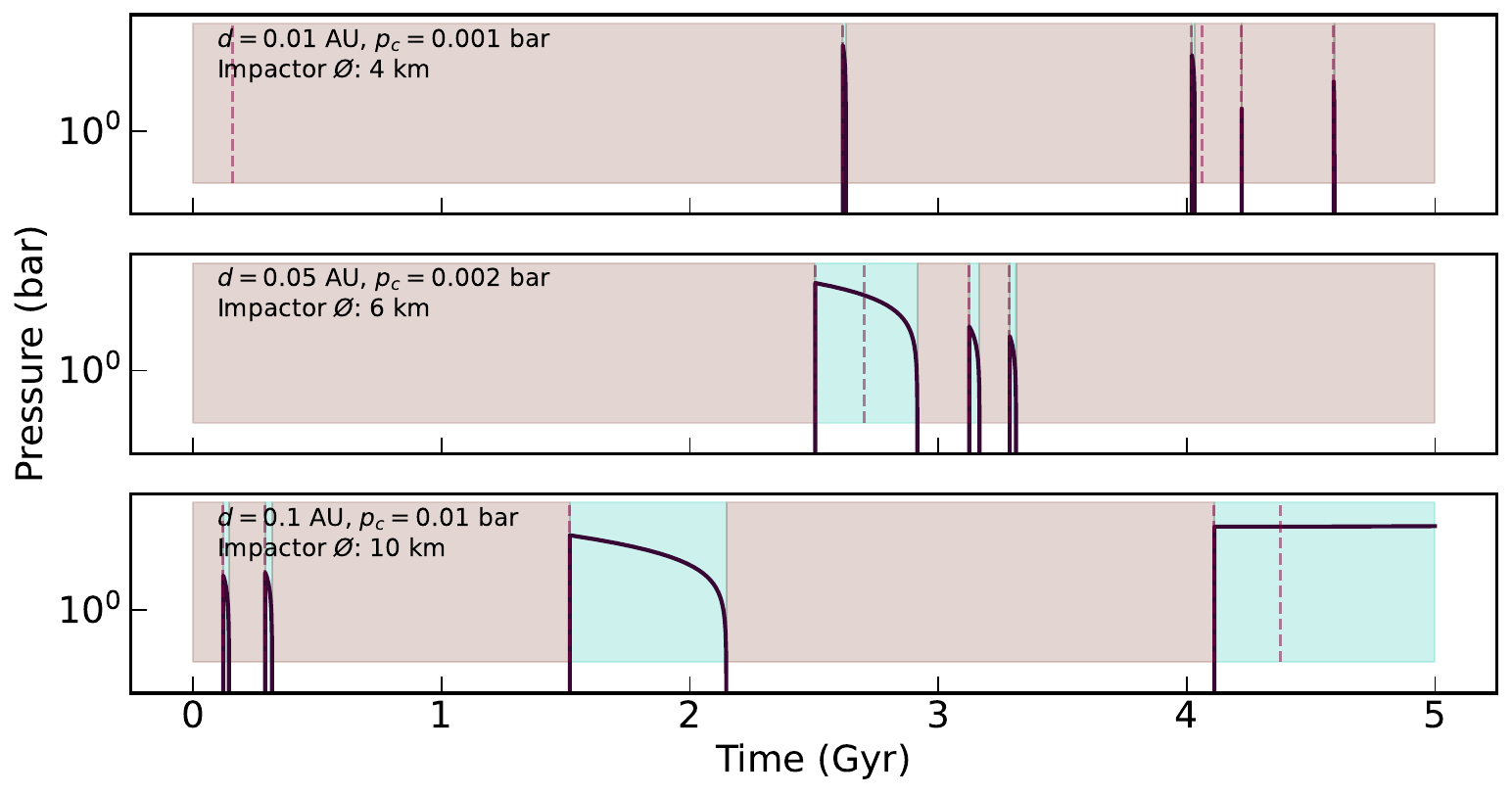}
    \caption{Evolution of an Earth-like planet around an M0-type star at different orbital distances, with inflated (turquoise) and collapsed (brown) states. The atmospheric evolution is uniquely determined by escape ($\eta = 0.01$), outgassing (current Earth CO$_2$ outgassing rates), collapse, and reinflation triggered by impacts (signified by the red dashed lines, impact rate $10^{-9}~\rm{yr}^{-1}$), as shown in Equation \eqref{eq:evolution}.}
    \label{fig:3earths}
\end{figure*}

\section{Results}\label{sec:results}
\subsection{Earth-like planet around an M-dwarf}
Figure \ref{fig:3earths} illustrates the basic mechanism for an Earth-like planet around an M dwarf with a bolometric luminosity of $0.072~L_{\odot}$, radius of $0.62~R_{\odot}$ and saturation time of $t_{sat} = 1~\rm{Gyr}$ (representative of an early M-type star). The 1 Earth-radius, 1 Earth-mass planet is placed at orbital distances of $d = 0.01, 0.05, \rm{and}~0.1~\rm{AU}$ respectively, and has a fixed CO$_2$ outgassing rate corresponding to modern Earth (see Section \ref{subsec:2state}). In this example, the impact rate for relevant impactors is fixed at $\tau = 10^{-9}~\rm{yr}^{-1}$, that is, on average, one impact every billion years. 

Each simulation starts without an atmosphere, under the assumption that the primary H/He envelope has been fully lost. This choice allows us to evaluate whether secondary atmospheres could exist around rocky exoplanets orbiting low-mass stars regardless of whether the primary atmosphere survived the high-XUV pre-main sequence phase. Observations also suggest that such planets do not typically retain significant H/He atmospheres \citep{Coy2025}. 

We find that impact-driven atmospheric regeneration produces longer-lasting atmospheres at low XUV irradiation - that is, farther from the star, and after the stellar saturation phase. At $0.01~\rm{AU}$, transient atmospheres typically collapse after a few million years, and sometimes even within one single time-step ($\leq10^{6}~\rm{yrs}$). At $d=0.05~\rm{AU}$, atmospheres can last 10-100 million years at a time once outside stellar saturation phase. At $d=0.1~\rm{AU}$, the atmospheres are much longer lived and can even subsist for a few billion years.

Another factor influencing the survival timescale of a transient atmosphere is the duration of the collapsed state preceding it. If an impact occurs shortly after atmospheric collapse, there are fewer volatiles injected into the atmosphere. Conversely, if the system remained collapsed for an extended period, accumulating volatiles on the nightside, the impact-generated atmosphere will be more massive and thus longer lived. The implications of this are further discussed in Section~\ref{subsec:disc_role}.

\subsection{Application to the Rocky Worlds targets}
We now apply our model to three rocky targets from the JWST DDT \textit{Rocky Worlds} programme \citep{Redfield2024}: LTT~1445~Ab \citep{Winters2019}, LTT~1445~Ac \citep{Winters2022}, and GJ~3929~b \citep{Kemmer2022}.
Instead of focusing on a static, final state of the evolution, we compute the fraction of time each planet spends with an inflated atmosphere. This approach accounts for the presence of transient atmospheres, such as the ones generated by impacts.

\begin{table}
\caption{Parameters used in the simulation for the three planets considered.}
\label{tab:plaprop}
\centering
\begin{tabular}{lllll}
\toprule
Planet              & T$_{\rm{eq}}$ (K) & p$_c$ (mbar) & D (km) & t$_{sat}$ (Gyrs) \\ \midrule
GJ~3929~b   & 568 & 5.03 & 6 & 1.1 \\
LTT~1445~Ac & 516 & 5.57 & 8 & 1.9 \\
LTT~1445~Ab & 431 & 18.28 & 10 & 1.9 \\ \bottomrule
\end{tabular}
\end{table}

GJ~3929~b ($R_p=1.09~R_{\oplus}$, $M_p=1.75~M_{\oplus}$; \citealt{Beard2022}) orbits an M3.5V star at $d=0.0252~\rm{AU}$. LTT~1445~Ac ($R_p=1.07~R_{\oplus}$, $M_p=1.35~M_{\oplus}$) and LTT~1445~Ab ($R_p=1.34~R_{\oplus}$, $M_p=2.73~M_{\oplus}$) orbit an M4 star at $d=0.0266~\rm{AU}$ and $d=0.0381~\rm{AU}$ respectively \citep{Pass2023}. Critical pressures $p_c$ for collapse and the minimum impactor diameter $D$ (Table~\ref{tab:plaprop}) are computed using Equations \ref{eq:ekin}-\ref{eq:diameter}, assuming $\rho_{imp} = 3.0~\rm{g~cm}^{-3}$. The saturation times are derived using the relationship with stellar mass for mid-to-late M dwarfs outlined in \citet{Pass2025}.

We perform $50,000$ Monte Carlo simulations, sampling impact rates uniformly in log scale from $10^{-10}$ to $10^{-6}~\rm{yr}^{-1}$ (0.1-1000 impacts per Gyr), and the CO$_2$ outgassing rates uniformly from 0.1-10 times modern Earth values. We compute inflation percentages over $2.2-12~\rm{Gyr}$ (see Appendix~\ref{app:stellar_ages} for a justification).

\begin{figure*}
    \centering
    \includegraphics[width=\linewidth]{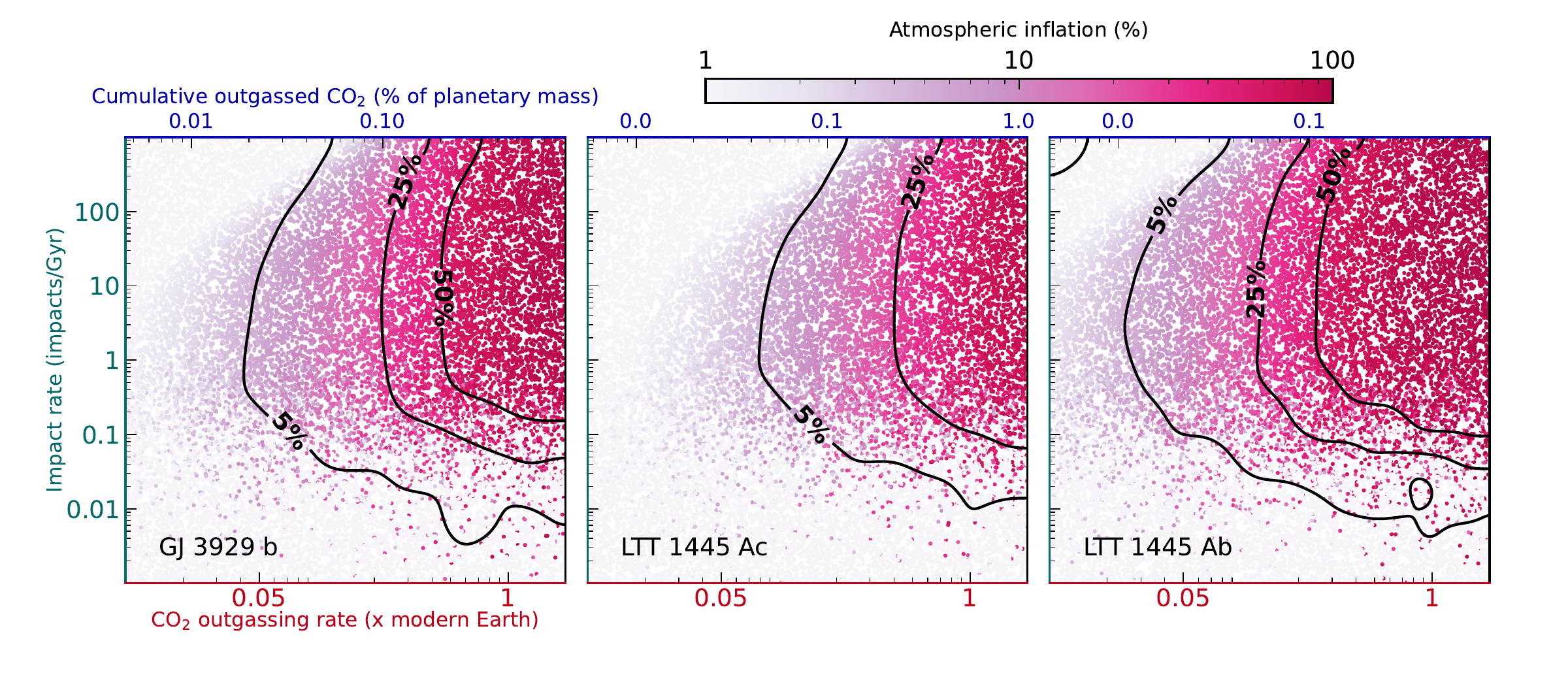}
    \caption{Fraction of time spent with transient CO$_2$ atmospheres generated by impacts between 2.2 and 12~Gyr of evolution.}
    \label{fig:rockyworlds}
\end{figure*}

Figure~\ref{fig:rockyworlds} shows the results of the Monte Carlo simulations, with each dot coloured by the percentage of time the planet spent in an inflated state for different CO$_2$-outgassing and impact rates. We show equivalent CO$_2$ inventories for each outgassing rate, which represent upper limits since we do not model carbon recycling. The contour lines for 5, 25, and 50$\%$ are shown in black.

The plots reveal an optimal range of impact rate for atmospheric regeneration. Below $0.1$ impacts/Gyr, insufficient impacts occur over the $\sim$10~Gyr simulation. However, a single, well-timed impact (i.e. after the saturation period and a longer collapsed phase) is able to generate a long-lasting atmosphere, as shown by the isolated dark pink dots on the bottom of the plots. Above $10-100$ impacts/Gyr, depending on outgassing rate, the system lacks sufficient time between impacts to accumulate volatiles on the nightside, reducing regeneration efficiency. This trend is consistent across planets.

Despite its higher equilibrium temperature, GJ~3929~b is more successful at atmospheric regeneration than LTT~1445~Ac. This is because GJ~3929 exits saturation phase before the later-type M dwarf LTT~1445~A. At $t=2.2~\rm{Gyr}$, GJ~3929~b already receives less XUV irradiation than LTT~1445~Ac due to the rapid decay in Equation~\eqref{eq:fxuv}, leading to longer-lived transient atmospheres. LTT~1445~Ab, which receives the least XUV irradiation, is especially favoured by this mechanism, with regions of the parameter space yielding $>50\%$ of transient atmospheric coverage.

Assuming current Earth outgassing and impact rates, GJ~3929~b, LTT~1445~Ac, and LTT~1445~Ab could spend around $65\%$, $45\%$, and $75\%$ of their lifetime respectively with a transient CO$_2$ atmosphere. For outgassing rates ten time lower, these fractions are around $10\%$, $5\%$, and $20\%$ respectively.

\section{Discussion}\label{sec:discussion}
\subsection{The protective role of atmospheric collapse}\label{subsec:disc_role}
A key insight from our simulations is that atmospheric collapse can shield volatiles from escape. While high XUV fluxes are generally assumed to irreversibly remove atmospheres, our results show that, paradoxically, they can help preserve volatiles by triggering collapse. Even if primordial H/He atmospheres are stripped early, volatiles left behind may accumulate as ice on the nightside, where they remain protected from XUV-driven loss.

For impact-driven regeneration to be effective, impacts must coexist with collapse. If significant impacts occur too frequently ($>100$/Gyr in our simulations), the reservoir is constantly depleted and transient atmospheres become short-lived. However, if collapsed periods are long enough to allow for build-up, one impact suffices to generate a large transient.

\subsection{Sink and source models for escape and outgassing}\label{subsec:disc_escout}
The atmospheric escape formula in Equation~\eqref{eq:escape} is intentionally simple, with much of the complexity absorbed into the efficiency factor $\eta$ \citep[see e.g.][for a discussion]{Ji2025}. For pure hydrogen, literature values range from $0.1$ to $0.6$ \citep{Lopez2012, Wordsworth2018, Erkaev2016}. CO$_2$ is more resistant to escape due to its higher molecular mass and ability to re-radiate effectively in the infrared \citep{WordsworthKreidberg2022}. Direct models of carbon thermal escape fluxes \citep{Tian2009} and stellar-wind driven ionic escape \citep{Dong2018} predict atmospheric removal rates lower than our assumptions (see Figure~\ref{fig:co2fluxes}). Our model is deliberately conservative, adopting CO$_2$ loss rates modestly higher than \cite{Tian2009} to account for uncertainties in escape processes. In contrast, stellar-wind driven ionic escape rates are orders of magnitude lower than our thermal escape estimates.

The impacts modelled in this work can also erode atmospheres, but they remove at most $h/2R$ of the total atmosphere \citep{Schlichting2015, Schlichting2018}, where $h$ is the scale height and $R$ the radius of the planet. For terrestrial planets with non H/He-dominated atmospheres (i.e. small scale height $h$), this corresponds to a small fraction (e.g. $0.07\%$ for Earth), and can be considered as being folded into the atmospheric escape sink term. 

Just as the escape is a sink term, outgassing serves as a source term. Our outgassing rates yield total CO$_2$ inventories ($0.01\%$ to $1\%$ of the planetary mass) comparable to initial volatile budgets assumed in other studies \citep[e.g.][]{Ji2025}. Actual outgassing likely varies with planetary evolution, tectonics, and thermal history. A thorough modelling of these processes could provide more realistic time-dependent outgassing rates, which would be a valuable extension for future work.

Incoming impactors can also deliver significant portions of volatiles, especially via smaller impactors not modelled in our work. For the outermost TRAPPIST-1 planets, \cite{Kral2018} estimate that the volatile mass accreted through exocomet and asteroid impacts could exceed that which is lost. This contribution, while not explicitly modelled here, could be considered to be incorporated into the outgassing source term.

\subsection{Impact rates}\label{subsec:disc_impactrates}
Estimating impact rates for exoplanetary systems remains highly uncertain, depending on factors like the presence and structure of debris belts, and the planetary system architecture. For our mechanism, the critical parameter is the minimum diameter $D$ necessary to trigger reinflation ($\sim 5-10~\rm{km}$ for the planets considered). Finally, the geometric distribution of volatile reservoirs should also be considered. An impactor has a higher probability of striking ice for nightside-wide ice sheets compared to polar caps.

Impactor populations typically follow a power-law size distribution, $N(D) = D^{-\gamma}$, where $\gamma \sim 3.5$ \citep{Schlichting2015, Kral2018}. However, translating this to absolute rates is challenging. Even for Earth, estimates for $\sim 5-10~\rm{km}$ sized impactors span an order of magnitude, from $10-100$ to $1-10$ impacts/Gyr). 

Our exploration of a broad plausible range from $<0.1$ to $1000$ impacts/Gyr allows to encompass these uncertainties while revealing that the mechanism effectively operates most effectively at intermediate rates ($1-100$ impacts/Gyr). Future constraints on exoplanetary debris disks and impact rate estimates will be crucial for refining these estimates.

\subsection{Mixed and H$_2$O-dominated atmospheres}

Although our model focuses on CO$_2$, the mechanism is general. Volatiles like H$_2$O, N$_2$ and O$_2$ could also participate, provided the nightside reaches temperatures low enough for them to condense. The case of H$_2$O is particularly interesting, as the molecule is readily photodissociated into H- and O- bearing products, including H$_2$ and O$_2$. H$_2$ is also an important expected by-product of impact degassing \citep{Haberle2019}. Mixed atmospheres would require careful calculations of the individual condensation temperatures based on the partial pressure of a given species \citep{HengKopparla2012}, a fractionated escape model \citep[e.g.,][]{Cherubim2024}, and incorporation of atmospheric chemistry and possibly surface-atmosphere interactions. We leave the treatment of this interesting problem to future work.

\subsection{Implications for observations}
Our work proposes a shift from the traditional atmosphere evolution picture wherein planets evolve smoothly from initial conditions to a final observable state. Instead, atmospheres on terrestrial exoplanets may be transient, governed not only by bulk properties, but also by episodic regeneration mechanisms. This dynamic view is observationally important, as it suggests detection rates may reflect atmospheric persistence rather than evolutionary endpoints.

The percentage of inflation provides a probabilistic framework to explore the parameter space. If a planet spends $1-10~\%$ of its time with an atmosphere, we should expect a corresponding success rate in detecting it. For LTT~1445~Ab, this fraction may even exceed $50~\%$, demonstrating that impact-driven atmospheric regeneration represents a viable pathway for maintaining detectable atmospheres around rocky exoplanets. 

The first observations for GJ~3929~b suggest it lacks a present-day atmosphere \citep{Xue2025}, consistent with our model predictions for CO$_2$ outgassing rates at or below Earth's current levels. This is plausible for a planet without significant ongoing geologic activity. Another possible explanation is the scarcity ($\leq 0.1$ impacts/Gyr) of impactors with diameters $\geq 6~\rm{km}$. 
Transient atmospheres may nonetheless have existed and could have left surface signatures (e.g., oxidation) detectable through dayside emission spectroscopy.

Crucially, equilibrium temperature alone may be an insufficient predictor for atmospheric presence. While useful for the retention of primary H/He atmospheres, our results show that transient secondary atmospheres are more sensitive to stellar type (via saturation phase duration), system age, and the geologic history.

The stochastic nature of transient atmospheres highlights the importance of surveying targets on both sides of the standard `cosmic shoreline' \citep{ZahnleCatling2017}. As observational datasets grow, the presence (or absence) of atmospheres around rocky exoplanets will offer new constraints on the geological processes governing the long-term evolution of secondary atmospheres.

\section{Conclusion}\label{sec:conclusion}
We have demonstrated that meteorite impacts can regenerate transient atmospheres on terrestrial planets orbiting low-mass stars through a simple energy balance mechanism. Our key findings are:
\begin{enumerate}
    \item Atmospheric collapse, though typically seen as detrimental to the survival of atmospheres around tidally locked rocky exoplanets, plays a protective role for volatiles by shielding them from atmospheric escape.
    \item This nightside ice reservoir can store substantial quantities of volatiles that remain available for atmospheric regeneration via impact vaporisation.
    \item Moderate-sized impactors with diameters ranging around $5-10~\rm{km}$ provide sufficient energy to vaporise large quantities of condensed CO$_2$ and thereby generate long-lasting transient atmospheres.
    \item The ideal impact rate for this mechanism to work is around $1-100~$ impacts/Gyr. Too frequent impacts hinder the replenishment of the volatile reservoir.
    \item We offer a probabilistic metric to assess the presence of atmospheres around rocky exoplanets to take into account the dynamic nature of secondary atmospheres. Under this metric, rocky planets around M-dwarfs could retain detectable CO$_2$ atmosphere for about $1-45~\%$ of their lifetime under plausible conditions.
\end{enumerate}

Our results suggest our traditional picture of a monotonic atmospheric evolution for rocky exoplanets may be incomplete. Instead, these planets may experience episodic atmospheric cycles driven by the interplay between stellar evolution, meteorite bombardment and volatile outgassing and recycling processes.

In this picture, atmospheric non-detections may be episodically collapsed atmospheres rather than permanently stripped ones. It is therefore important for JWST and follow-up missions to keep probing terrestrial targets across the cosmic shoreline. Large sample sizes will reveal probabilistic trends, which can in turn shed light on the processes dominating transient atmospheric regeneration. This framework also suggests that continued observations improve the likelihood of eventually detecting an atmosphere on a rocky M-dwarf planet.

\begin{acknowledgments}
P.C.A. acknowledges support from the Carlsberg Foundation, grant CF22-1254. R.W. acknowledges support from Leverhulme Center for Life in the Universe grant G119167, LBAG/312.
\end{acknowledgments}

\begin{contribution}
P.A. led the modelling, analysis, and manuscript preparation. R.W. conceived and supervised the project. M. H. and D.B. contributed expertise on impacts and atmospheric loss. L.A.B. provided overall supervision and manuscript feedback.
\end{contribution}

%



\appendix
\section{CO$_2$ escape rates comparison} \label{app:co2esc}
\begin{figure}
    \centering
    \includegraphics[width=0.6\linewidth]{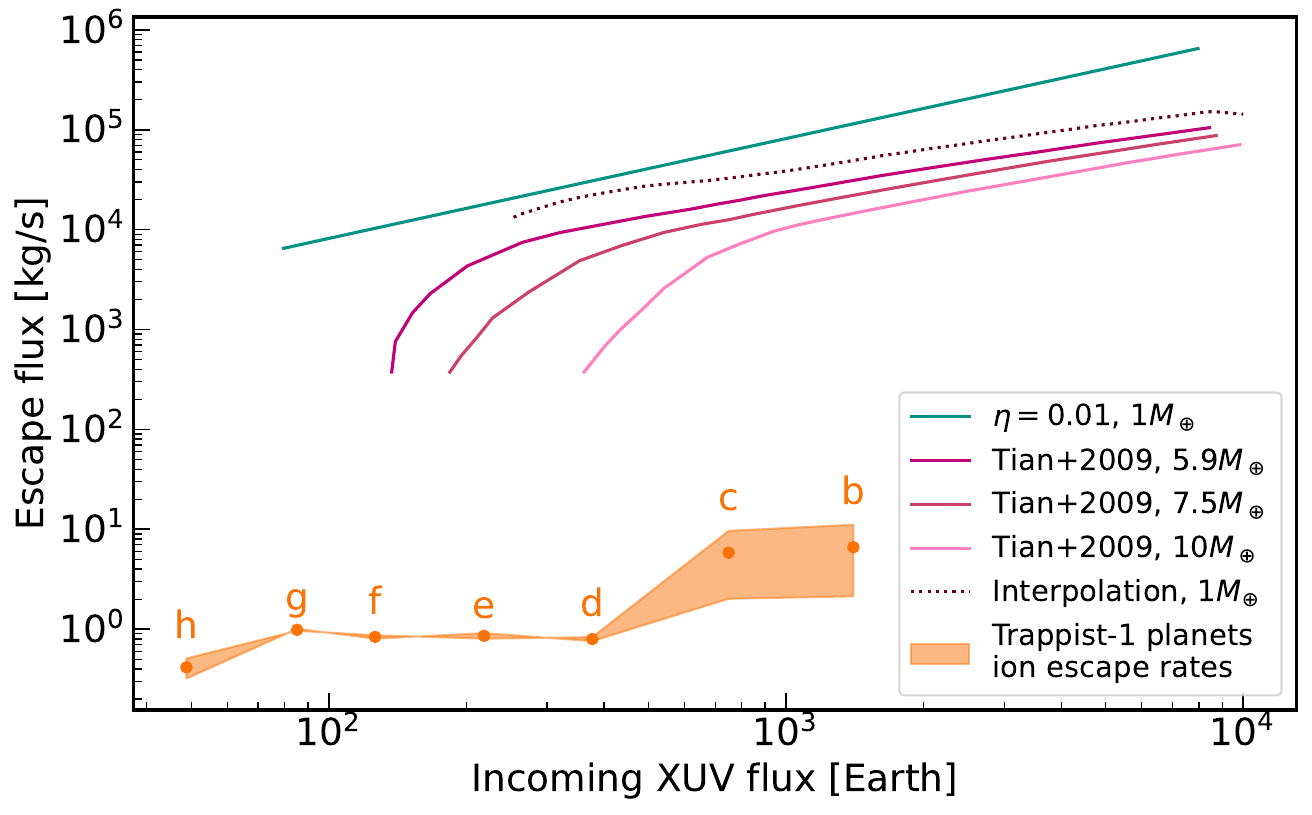}
    \caption{Comparison of CO$_2$ escape fluxes as modeled using energy-limited escape with an efficiency $\eta=0.1$ (green line), carbon thermal escape fluxes for different super-Earths \citep[][pink lines]{Tian2009}, and stellar-wind driven ion escape for the TRAPPIST-1 planets \citep[][orange shaded area]{Dong2018}.}
    \label{fig:co2fluxes}
\end{figure}

Figure~\ref{fig:co2fluxes} shows a comparison of different escape mass fluxes for CO$_2$ using our nominal model (energy-limited escape with $\eta=0.01$) on an Earth-size, Earth-mass planet, the \citet{Tian2009} carbon thermal escape models, and the stellar wind-induced atmospheric loss of CO$_2^+$ ions from \citet{Dong2018} for the different TRAPPIST-1 planets. In this last case, we place the data points on the x-axis using the XUV fluxes each planet receives.

\section{Stellar ages}\label{app:stellar_ages}
While the age of M dwarfs is notoriously difficult to estimate, their rotation rates can serve as proxies \citep{Engle2023}. For GJ~3929, \citet{Kemmer2022} report a stellar rotation period of $P_* = 122\pm16~\rm{days}$, placing the star the star in an age bin of $12.9\pm3.5~\rm{Gyr}$ according to \citet{Medina2022}. This is consistent with the value of  $7.1^{+4.1}_{-4.9}~\rm{Gyr}$ reported by \citet{Beard2022}. Similarly, the LTT~1445 system is thought to be old, with a reported rotation period of $P_* = 85\pm22$ \citep{Winters2022} for LTT~1445~A, placing it between the middle and older bins of \citet{Medina2022}, that is $5.6\pm2.7~\rm{Gyr}$ and $12.9\pm3.5~\rm{Gyr}$. \citet{DiamondLowe2024} also report a conservative lower age limit of $2.2~\rm{Gyr}$. 

In the absence of more precise constraints for the age of these two systems, and in an effort to get statistically meaningful but comparable results, we chose the common lower limit of $2.2~\rm{Gyr}$ for the three targets, and compute the inflation percentages over a simulation up to $12~\rm{Gyr}$. 





\bibliographystyle{aasjournalv7}
\bibliography{references}

\begin{thebibliography}{}
\expandafter\ifx\csname natexlab\endcsname\relax\def\natexlab#1{#1}\fi
\providecommand{\url}[1]{\href{#1}{#1}}
\providecommand{\dodoi}[1]{doi:~\href{http://doi.org/#1}{\nolinkurl{#1}}}
\providecommand{\doeprint}[1]{\href{http://ascl.net/#1}{\nolinkurl{http://ascl.net/#1}}}
\providecommand{\doarXiv}[1]{\href{https://arxiv.org/abs/#1}{\nolinkurl{https://arxiv.org/abs/#1}}}

\bibitem[{T.~J. {Ahrens} \& J.~D. {O'Keefe}(1972){Ahrens} \& {O'Keefe}}]{AhrensOKeefe1972}
{Ahrens}, T.~J., \& {O'Keefe}, J.~D. 1972, \bibinfo{title}{{Shock melting and vaporization of lunar rocks and minerals},} Moon, 4, 214, \dodoi{10.1007/BF00562927}

\bibitem[{L.~W. Alvarez {et~al.}(1980)Alvarez, Alvarez, Asaro, \& Michel}]{Alvarez1980}
Alvarez, L.~W., Alvarez, W., Asaro, F., \& Michel, H.~V. 1980, \bibinfo{title}{Extraterrestrial Cause for the Cretaceous-Tertiary Extinction,} Science, 208, 1095, \dodoi{10.1126/science.208.4448.1095}

\bibitem[{P.~C. {August}(2025){August}}]{August2025code}
{August}, P.~C. 2025, Model for atmospheric collapse and re-inflation through impacts, Zenodo, \dodoi{10.5281/zenodo.17821831}

\bibitem[{P.~C. {August} {et~al.}(2025){August}, {Buchhave}, {Diamond-Lowe}, {Mendon{\c{c}}a}, {Gressier}, {Rathcke}, {Allen}, {Fortune}, {Jones}, {Meier Vald{\'e}s}, {Demory}, {Espinoza}, {Fisher}, {Gibson}, {Heng}, {Hoeijmakers}, {Hooton}, {Kitzmann}, {Prinoth}, {Eastman}, \& {Barnes}}]{August2025}
{August}, P.~C., {Buchhave}, L.~A., {Diamond-Lowe}, H., {et~al.} 2025, \bibinfo{title}{{Hot Rocks Survey I: A possible shallow eclipse for LHS 1478 b},} \aap, 695, A171, \dodoi{10.1051/0004-6361/202452611}

\bibitem[{C.~J. {Barnhart} \& F. {Nimmo}(2011){Barnhart} \& {Nimmo}}]{Barnhart2011}
{Barnhart}, C.~J., \& {Nimmo}, F. 2011, \bibinfo{title}{{Role of impact excavation in distributing clays over Noachian surfaces},} Journal of Geophysical Research (Planets), 116, E01009, \dodoi{10.1029/2010JE003629}

\bibitem[{C. {Beard} {et~al.}(2022){Beard}, {Robertson}, {Kanodia}, {Lubin}, {Ca{\~n}as}, {Gupta}, {Holcomb}, {Jones}, {Libby-Roberts}, {Lin}, {Mahadevan}, {Stef{\'a}nsson}, {Bender}, {Blake}, {Cochran}, {Endl}, {Everett}, {Ford}, {Fredrick}, {Halverson}, {Hebb}, {Li}, {Logsdon}, {Luhn}, {McElwain}, {Metcalf}, {Ninan}, {Rajagopal}, {Roy}, {Schutte}, {Schwab}, {Terrien}, {Wisniewski}, \& {Wright}}]{Beard2022}
{Beard}, C., {Robertson}, P., {Kanodia}, S., {et~al.} 2022, \bibinfo{title}{{GJ 3929: High-precision Photometric and Doppler Characterization of an Exo-Venus and Its Hot, Mini-Neptune-mass Companion},} \apj, 936, 55, \dodoi{10.3847/1538-4357/ac8480}

\bibitem[{A. {Bello-Arufe} {et~al.}(2025){Bello-Arufe}, {Damiano}, {Bennett}, {Hu}, {Welbanks}, {MacDonald}, {Seligman}, {Sing}, {Tokadjian}, {Oza}, \& {Yang}}]{BelloArufe2025}
{Bello-Arufe}, A., {Damiano}, M., {Bennett}, K.~A., {et~al.} 2025, \bibinfo{title}{{Evidence for a Volcanic Atmosphere on the Sub-Earth L 98-59 b},} \apjl, 980, L26, \dodoi{10.3847/2041-8213/adaf22}

\bibitem[{C. {Cherubim} {et~al.}(2024){Cherubim}, {Wordsworth}, {Hu}, \& {Shkolnik}}]{Cherubim2024}
{Cherubim}, C., {Wordsworth}, R., {Hu}, R., \& {Shkolnik}, E. 2024, \bibinfo{title}{{Strong Fractionation of Deuterium and Helium in Sub-Neptune Atmospheres along the Radius Valley},} \apj, 967, 139, \dodoi{10.3847/1538-4357/ad3e77}

\bibitem[{F.~J. {Ciesla} {et~al.}(2015){Ciesla}, {Mulders}, {Pascucci}, \& {Apai}}]{Ciesla2015}
{Ciesla}, F.~J., {Mulders}, G.~D., {Pascucci}, I., \& {Apai}, D. 2015, \bibinfo{title}{{Volatile Delivery to Planets from Water-rich Planetesimals around Low Mass Stars},} \apj, 804, 9, \dodoi{10.1088/0004-637X/804/1/9}

\bibitem[{B.~P. {Coy} {et~al.}(2025){Coy}, {Ih}, {Kite}, {Koll}, {Tenthoff}, {Bean}, {Weiner Mansfield}, {Zhang}, {Xue}, {Kempton}, {Wohlfarth}, {Hu}, {Lyu}, \& {W{\"o}hler}}]{Coy2025}
{Coy}, B.~P., {Ih}, J., {Kite}, E.~S., {et~al.} 2025, \bibinfo{title}{{Population-level Hypothesis Testing with Rocky Planet Emission Data: A Tentative Trend in the Brightness Temperatures of M-Earths},} \apj, 987, 22, \dodoi{10.3847/1538-4357/add3f7}

\bibitem[{H. {Diamond-Lowe} {et~al.}(2023){Diamond-Lowe}, {Mendonca}, {Akin}, {Allen}, {Baungaard}, {Borsato}, {Buchhave}, {Burgasser}, {Demory}, {Espinoza}, {Fisher}, {Fortune}, {Gibson}, {Gressier}, {Guzman Mesa}, {Heng}, {Hoeijmakers}, {Hooton}, {Jones}, {Kitzmann}, {Lueber}, {Meier Valdes}, {Prinoth}, {Rathcke}, \& {Tian}}]{DiamondLowe2023prop}
{Diamond-Lowe}, H., {Mendonca}, J.~M., {Akin}, C.~J., {et~al.} 2023, {The Hot Rocks Survey: Testing 9 Irradiated Terrestrial Exoplanets for Atmospheres},, JWST Proposal. Cycle 2, ID. \#3730

\bibitem[{H. {Diamond-Lowe} {et~al.}(2024){Diamond-Lowe}, {King}, {Youngblood}, {Brown}, {Howard}, {Winters}, {Wilson}, {France}, {Mendon{\c{c}}a}, {Buchhave}, {Corrales}, {Kreidberg}, {Medina}, {Bean}, {Berta-Thompson}, {Evans-Soma}, {Froning}, {Duvvuri}, {Kempton}, {Miguel}, {Pineda}, \& {Schneider}}]{DiamondLowe2024}
{Diamond-Lowe}, H., {King}, G.~W., {Youngblood}, A., {et~al.} 2024, \bibinfo{title}{{High-energy spectra of LTT 1445A and GJ 486 reveal flares and activity},} \aap, 689, A48, \dodoi{10.1051/0004-6361/202450107}

\bibitem[{L.~N.~R. do~Amaral {et~al.}(2022)do~Amaral, Barnes, Segura, \& Luger}]{doAmaral2022}
do~Amaral, L. N.~R., Barnes, R., Segura, A., \& Luger, R. 2022, \bibinfo{title}{The Contribution of M-dwarf Flares to the Thermal Escape of Potentially Habitable Planet Atmospheres,} The Astrophysical Journal, 928, 12, \dodoi{10.3847/1538-4357/ac53af}

\bibitem[{C. {Dong} {et~al.}(2018){Dong}, {Jin}, {Lingam}, {Airapetian}, {Ma}, \& {van der Holst}}]{Dong2018}
{Dong}, C., {Jin}, M., {Lingam}, M., {et~al.} 2018, \bibinfo{title}{{Atmospheric escape from the TRAPPIST-1 planets and implications for habitability},} Proceedings of the National Academy of Science, 115, 260, \dodoi{10.1073/pnas.1708010115}

\bibitem[{S.~G. {Engle} \& E.~F. {Guinan}(2023){Engle} \& {Guinan}}]{Engle2023}
{Engle}, S.~G., \& {Guinan}, E.~F. 2023, \bibinfo{title}{{Living with a Red Dwarf: The Rotation-Age Relationships of M Dwarfs},} \apjl, 954, L50, \dodoi{10.3847/2041-8213/acf472}

\bibitem[{N.~V. {Erkaev} {et~al.}(2016){Erkaev}, {Lammer}, {Odert}, {Kislyakova}, {Johnstone}, {G{\"u}del}, \& {Khodachenko}}]{Erkaev2016}
{Erkaev}, N.~V., {Lammer}, H., {Odert}, P., {et~al.} 2016, \bibinfo{title}{{EUV-driven mass-loss of protoplanetary cores with hydrogen-dominated atmospheres: the influences of ionization and orbital distance},} \mnras, 460, 1300, \dodoi{10.1093/mnras/stw935}

\bibitem[{F.~P. {Fanale} {et~al.}(1982){Fanale}, {Salvail}, {Banerdt}, \& {Saunders}}]{Fanale1982}
{Fanale}, F.~P., {Salvail}, J.~R., {Banerdt}, W.~B., \& {Saunders}, R.~S. 1982, \bibinfo{title}{{Mars: The regolith-atmosphere-cap system and climate change},} \icarus, 50, 381, \dodoi{10.1016/0019-1035(82)90131-2}

\bibitem[{F. {Forget} {et~al.}(2013){Forget}, {Wordsworth}, {Millour}, {Madeleine}, {Kerber}, {Leconte}, {Marcq}, \& {Haberle}}]{Forget2013}
{Forget}, F., {Wordsworth}, R., {Millour}, E., {et~al.} 2013, \bibinfo{title}{{3D modelling of the early martian climate under a denser CO$_{2}$ atmosphere: Temperatures and CO$_{2}$ ice clouds},} \icarus, 222, 81, \dodoi{10.1016/j.icarus.2012.10.019}

\bibitem[{M. {Fortune} {et~al.}(2025){Fortune}, {Gibson}, {Diamond-Lowe}, {Mendon{\c{c}}a}, {Gressier}, {Kitzmann}, {Allen}, {August}, {Ih}, {Meier Vald{\'e}s}, {Zgraggen}, {Buchhave}, {Demory}, {Espinoza}, {Heng}, {Jones}, \& {Rathcke}}]{Fortune2025}
{Fortune}, M., {Gibson}, N.~P., {Diamond-Lowe}, H., {et~al.} 2025, \bibinfo{title}{{Hot Rocks Survey III: A deep eclipse for LHS 1140c and a new Gaussian process method to account for correlated noise in individual pixels},} arXiv e-prints, arXiv:2505.22186, \dodoi{10.48550/arXiv.2505.22186}

\bibitem[{R.~M. {Haberle} {et~al.}(2019){Haberle}, {Zahnle}, {Barlow}, \& {Steakley}}]{Haberle2019}
{Haberle}, R.~M., {Zahnle}, K., {Barlow}, N.~G., \& {Steakley}, K.~E. 2019, \bibinfo{title}{{Impact Degassing of H$_{2}$ on Early Mars and its Effect on the Climate System},} \grl, 46, 13,355, \dodoi{10.1029/2019GL084733}

\bibitem[{W.~K. {Hartmann} \& G. {Neukum}(2001){Hartmann} \& {Neukum}}]{HartmannNeukum2001}
{Hartmann}, W.~K., \& {Neukum}, G. 2001, \bibinfo{title}{{Cratering Chronology and the Evolution of Mars},} \ssr, 96, 165, \dodoi{10.1023/A:1011945222010}

\bibitem[{K. {Heng} \& P. {Kopparla}(2012){Heng} \& {Kopparla}}]{HengKopparla2012}
{Heng}, K., \& {Kopparla}, P. 2012, \bibinfo{title}{{On the Stability of Super-Earth Atmospheres},} \apj, 754, 60, \dodoi{10.1088/0004-637X/754/1/60}

\bibitem[{R. {Hu} {et~al.}(2024){Hu}, {Bello-Arufe}, {Zhang}, {Paragas}, {Zilinskas}, {van Buchem}, {Bess}, {Patel}, {Ito}, {Damiano}, {Scheucher}, {Oza}, {Knutson}, {Miguel}, {Dragomir}, {Brandeker}, \& {Demory}}]{Hu2024}
{Hu}, R., {Bello-Arufe}, A., {Zhang}, M., {et~al.} 2024, \bibinfo{title}{{A secondary atmosphere on the rocky exoplanet 55 Cancri e},} \nat, 630, 609, \dodoi{10.1038/s41586-024-07432-x}

\bibitem[{X. {Ji} {et~al.}(2025){Ji}, {Chatterjee}, {Park Coy}, \& {Kite}}]{Ji2025}
{Ji}, X., {Chatterjee}, R.~D., {Park Coy}, B., \& {Kite}, E.~S. 2025, \bibinfo{title}{{The Cosmic Shoreline Revisited: A Metric for Atmospheric Retention Informed by Hydrodynamic Escape},} arXiv e-prints, arXiv:2504.19872, \dodoi{10.48550/arXiv.2504.19872}

\bibitem[{M.~M. {Joshi} {et~al.}(1997){Joshi}, {Haberle}, \& {Reynolds}}]{Joshi1997}
{Joshi}, M.~M., {Haberle}, R.~M., \& {Reynolds}, R.~T. 1997, \bibinfo{title}{{Simulations of the Atmospheres of Synchronously Rotating Terrestrial Planets Orbiting M Dwarfs: Conditions for Atmospheric Collapse and the Implications for Habitability},} \icarus, 129, 450, \dodoi{10.1006/icar.1997.5793}

\bibitem[{J.~F. {Kasting} {et~al.}(1993){Kasting}, {Whitmire}, \& {Reynolds}}]{Kasting1993}
{Kasting}, J.~F., {Whitmire}, D.~P., \& {Reynolds}, R.~T. 1993, \bibinfo{title}{{Habitable Zones around Main Sequence Stars},} \icarus, 101, 108, \dodoi{10.1006/icar.1993.1010}

\bibitem[{J. {Kemmer} {et~al.}(2022){Kemmer}, {Dreizler}, {Kossakowski}, {Stock}, {Quirrenbach}, {Caballero}, {Amado}, {Collins}, {Espinoza}, {Herrero}, {Jenkins}, {Latham}, {Lillo-Box}, {Narita}, {Pall{\'e}}, {Reiners}, {Ribas}, {Ricker}, {Rodr{\'\i}guez}, {Seager}, {Vanderspek}, {Wells}, {Winn}, {Aceituno}, {B{\'e}jar}, {Barclay}, {Bluhm}, {Chaturvedi}, {Cifuentes}, {Collins}, {Cort{\'e}s-Contreras}, {Demory}, {Fausnaugh}, {Fukui}, {G{\'o}mez Maqueo Chew}, {Galad{\'\i}-Enr{\'\i}quez}, {Gan}, {Gillon}, {Golovin}, {Hatzes}, {Henning}, {Huang}, {Jeffers}, {Kaminski}, {Kunimoto}, {K{\"u}rster}, {L{\'o}pez-Gonz{\'a}lez}, {Lafarga}, {Luque}, {McCormac}, {Molaverdikhani}, {Montes}, {Morales}, {Passegger}, {Reffert}, {Sabin}, {Sch{\"o}fer}, {Schanche}, {Schlecker}, {Schroffenegger}, {Schwarz}, {Schweitzer}, {Sota}, {Tenenbaum}, {Trifonov}, {Vanaverbeke}, \& {Zechmeister}}]{Kemmer2022}
{Kemmer}, J., {Dreizler}, S., {Kossakowski}, D., {et~al.} 2022, \bibinfo{title}{{Discovery and mass measurement of the hot, transiting, Earth-sized planet, GJ 3929 b},} \aap, 659, A17, \dodoi{10.1051/0004-6361/202142653}

\bibitem[{D.~D.~B. {Koll} \& D.~S. {Abbot}(2016){Koll} \& {Abbot}}]{Koll2016}
{Koll}, D. D.~B., \& {Abbot}, D.~S. 2016, \bibinfo{title}{{Temperature Structure and Atmospheric Circulation of Dry Tidally Locked Rocky Exoplanets},} \apj, 825, 99, \dodoi{10.3847/0004-637X/825/2/99}

\bibitem[{Q. {Kral} {et~al.}(2018){Kral}, {Wyatt}, {Triaud}, {Marino}, {Th{\'e}bault}, \& {Shorttle}}]{Kral2018}
{Kral}, Q., {Wyatt}, M.~C., {Triaud}, A. H.~M.~J., {et~al.} 2018, \bibinfo{title}{{Cometary impactors on the TRAPPIST-1 planets can destroy all planetary atmospheres and rebuild secondary atmospheres on planets f, g, and h},} \mnras, 479, 2649, \dodoi{10.1093/mnras/sty1677}

\bibitem[{R.~G. {Kraus} {et~al.}(2011){Kraus}, {Senft}, \& {Stewart}}]{Kraus2011}
{Kraus}, R.~G., {Senft}, L.~E., \& {Stewart}, S.~T. 2011, \bibinfo{title}{{Impacts onto H $_{2}$O ice: Scaling laws for melting, vaporization, excavation, and final crater size},} \icarus, 214, 724, \dodoi{10.1016/j.icarus.2011.05.016}

\bibitem[{J. {Krissansen-Totton}(2023){Krissansen-Totton}}]{Krissansen2023}
{Krissansen-Totton}, J. 2023, \bibinfo{title}{{Implications of Atmospheric Nondetections for Trappist-1 Inner Planets on Atmospheric Retention Prospects for Outer Planets},} \apjl, 951, L39, \dodoi{10.3847/2041-8213/acdc26}

\bibitem[{M. {Le Feuvre} \& M.~A. {Wieczorek}(2011){Le Feuvre} \& {Wieczorek}}]{LeFeuvre2011}
{Le Feuvre}, M., \& {Wieczorek}, M.~A. 2011, \bibinfo{title}{{Nonuniform cratering of the Moon and a revised crater chronology of the inner Solar System},} \icarus, 214, 1, \dodoi{10.1016/j.icarus.2011.03.010}

\bibitem[{A. {Lecavelier des Etangs} {et~al.}(2022){Lecavelier des Etangs}, {Cros}, {H{\'e}brard}, {Martioli}, {Duquesnoy}, {Kenworthy}, {Kiefer}, {Lacour}, {Lagrange}, {Meunier}, \& {Vidal-Madjar}}]{Lecavelier2022}
{Lecavelier des Etangs}, A., {Cros}, L., {H{\'e}brard}, G., {et~al.} 2022, \bibinfo{title}{{Exocomets size distribution in the {\ensuremath{\beta}} Pictoris planetary system},} Scientific Reports, 12, 5855, \dodoi{10.1038/s41598-022-09021-2}

\bibitem[{E.~D. {Lopez} {et~al.}(2012){Lopez}, {Fortney}, \& {Miller}}]{Lopez2012}
{Lopez}, E.~D., {Fortney}, J.~J., \& {Miller}, N. 2012, \bibinfo{title}{{How Thermal Evolution and Mass-loss Sculpt Populations of Super-Earths and Sub-Neptunes: Application to the Kepler-11 System and Beyond},} \apj, 761, 59, \dodoi{10.1088/0004-637X/761/1/59}

\bibitem[{A.~A. {Medina} {et~al.}(2020){Medina}, {Winters}, {Irwin}, \& {Charbonneau}}]{Medina2020}
{Medina}, A.~A., {Winters}, J.~G., {Irwin}, J.~M., \& {Charbonneau}, D. 2020, \bibinfo{title}{{Flare Rates, Rotation Periods, and Spectroscopic Activity Indicators of a Volume-complete Sample of Mid- to Late-M Dwarfs within 15 pc},} \apj, 905, 107, \dodoi{10.3847/1538-4357/abc686}

\bibitem[{A.~A. {Medina} {et~al.}(2022){Medina}, {Winters}, {Irwin}, \& {Charbonneau}}]{Medina2022}
{Medina}, A.~A., {Winters}, J.~G., {Irwin}, J.~M., \& {Charbonneau}, D. 2022, \bibinfo{title}{{Galactic Kinematics and Observed Flare Rates of a Volume-complete Sample of Mid-to-late M Dwarfs: Constraints on the History of the Stellar Radiation Environment of Planets Orbiting Low-mass Stars},} \apj, 935, 104, \dodoi{10.3847/1538-4357/ac77f9}

\bibitem[{E.~A. {Meier Vald{\'e}s} {et~al.}(2025){Meier Vald{\'e}s}, {Demory}, {Diamond-Lowe}, {Mendon{\c{c}}a}, {August}, {Fortune}, {Allen}, {Kitzmann}, {Gressier}, {Hooton}, {Jones}, {Buchhave}, {Espinoza}, {Fisher}, {Gibson}, {Heng}, {Hoeijmakers}, {Prinoth}, {Rathcke}, \& {Eastman}}]{MeierValdes2025}
{Meier Vald{\'e}s}, E.~A., {Demory}, B.~O., {Diamond-Lowe}, H., {et~al.} 2025, \bibinfo{title}{{Hot Rocks Survey: II. The thermal emission of TOI-1468 b reveals a bare hot rock},} \aap, 698, A68, \dodoi{10.1051/0004-6361/202453449}

\bibitem[{H.~J. {Melosh}(1989){Melosh}}]{Melosh1989}
{Melosh}, H.~J. 1989, {Impact cratering : a geologic process}

\bibitem[{A. {Morbidelli} {et~al.}(2000){Morbidelli}, {Chambers}, {Lunine}, {Petit}, {Robert}, {Valsecchi}, \& {Cyr}}]{Morbidelli2000}
{Morbidelli}, A., {Chambers}, J., {Lunine}, J.~I., {et~al.} 2000, \bibinfo{title}{{Source regions and time scales for the delivery of water to Earth},} \maps, 35, 1309, \dodoi{10.1111/j.1945-5100.2000.tb01518.x}

\bibitem[{T. {Mukai} {et~al.}(1997){Mukai}, {Tanaka}, {Ishimoto}, \& {Nakamura}}]{Mukai1997}
{Mukai}, T., {Tanaka}, M., {Ishimoto}, H., \& {Nakamura}, R. 1997, \bibinfo{title}{{Temperature variations across craters in the polar regions of the Moon and Mercury},} Advances in Space Research, 19, 1497, \dodoi{10.1016/S0273-1177(97)00348-7}

\bibitem[{J.~D. {O'Keefe} \& T.~J. {Ahrens}(1977){O'Keefe} \& {Ahrens}}]{AhrensOKeefe1977}
{O'Keefe}, J.~D., \& {Ahrens}, T.~J. 1977, \bibinfo{title}{{Partitioning of Energy and the Degree of Melting and Vaporization in Planetary Impact Processes},} in Lunar and Planetary Science Conference, Vol.~8, Lunar and Planetary Science Conference, 741

\bibitem[{A.~M. {Palumbo} \& J.~W. {Head}(2018){Palumbo} \& {Head}}]{PalumboHead2018}
{Palumbo}, A.~M., \& {Head}, J.~W. 2018, \bibinfo{title}{{Impact cratering as a cause of climate change, surface alteration, and resurfacing during the early history of Mars},} \maps, 53, 687, \dodoi{10.1111/maps.13001}

\bibitem[{E.~K. {Pass} {et~al.}(2025){Pass}, {Charbonneau}, \& {Vanderburg}}]{Pass2025}
{Pass}, E.~K., {Charbonneau}, D., \& {Vanderburg}, A. 2025, \bibinfo{title}{{The Receding Cosmic Shoreline of Mid-to-late M Dwarfs: Measurements of Active Lifetimes Worsen Challenges for Atmosphere Retention by Rocky Exoplanets},} \apjl, 986, L3, \dodoi{10.3847/2041-8213/adda39}

\bibitem[{E.~K. {Pass} {et~al.}(2023){Pass}, {Winters}, {Charbonneau}, {Balkanski}, {Lewis}, {Lally}, {Bean}, {Cloutier}, \& {Eastman}}]{Pass2023}
{Pass}, E.~K., {Winters}, J.~G., {Charbonneau}, D., {et~al.} 2023, \bibinfo{title}{{HST/WFC3 Light Curve Supports a Terrestrial Composition for the Closest Exoplanet to Transit an M Dwarf},} \aj, 166, 171, \dodoi{10.3847/1538-3881/acf561}

\bibitem[{E. {Pierazzo} {et~al.}(1998){Pierazzo}, {Kring}, \& {Melosh}}]{Pierazzo1998}
{Pierazzo}, E., {Kring}, D.~A., \& {Melosh}, H.~J. 1998, \bibinfo{title}{{Hydrocode simulation of the Chicxulub impact event and the production of climatically active gases},} \jgr, 103, 28607, \dodoi{10.1029/98JE02496}

\bibitem[{S. {Redfield} {et~al.}(2024){Redfield}, {Batalha}, {Benneke}, {Biller}, {Espinoza}, {France}, {Konopacky}, {Kreidberg}, {Rauscher}, \& {Sing}}]{Redfield2024}
{Redfield}, S., {Batalha}, N., {Benneke}, B., {et~al.} 2024, \bibinfo{title}{{Report of the Working Group on Strategic Exoplanet Initiatives with HST and JWST},} arXiv e-prints, arXiv:2404.02932, \dodoi{10.48550/arXiv.2404.02932}

\bibitem[{I. {Ribas} {et~al.}(2005){Ribas}, {Guinan}, {G{\"u}del}, \& {Audard}}]{Ribas2005}
{Ribas}, I., {Guinan}, E.~F., {G{\"u}del}, M., \& {Audard}, M. 2005, \bibinfo{title}{{Evolution of the Solar Activity over Time and Effects on Planetary Atmospheres. I. High-Energy Irradiances (1-1700 {\r{A}})},} \apj, 622, 680, \dodoi{10.1086/427977}

\bibitem[{J.~A.~P. {Rodriguez} {et~al.}(2023){Rodriguez}, {Domingue}, {Travis}, {Kargel}, {Abramov}, {Zarroca}, {Banks}, {Weirich}, {Lopez}, {Castle}, {Jianguo}, \& {Chuang}}]{Rodriguez2023}
{Rodriguez}, J. A.~P., {Domingue}, D., {Travis}, B., {et~al.} 2023, \bibinfo{title}{{Mercury's Hidden Past: Revealing a Volatile-dominated Layer through Glacier-like Features and Chaotic Terrains},} \psj, 4, 219, \dodoi{10.3847/PSJ/acf219}

\bibitem[{H.~E. {Schlichting} \& S. {Mukhopadhyay}(2018){Schlichting} \& {Mukhopadhyay}}]{Schlichting2018}
{Schlichting}, H.~E., \& {Mukhopadhyay}, S. 2018, \bibinfo{title}{{Atmosphere Impact Losses},} \ssr, 214, 34, \dodoi{10.1007/s11214-018-0471-z}

\bibitem[{H.~E. {Schlichting} {et~al.}(2015){Schlichting}, {Sari}, \& {Yalinewich}}]{Schlichting2015}
{Schlichting}, H.~E., {Sari}, R., \& {Yalinewich}, A. 2015, \bibinfo{title}{{Atmospheric mass loss during planet formation: The importance of planetesimal impacts},} \icarus, 247, 81, \dodoi{10.1016/j.icarus.2014.09.053}

\bibitem[{T.~L. {Segura} {et~al.}(2008){Segura}, {Toon}, \& {Colaprete}}]{Segura2008}
{Segura}, T.~L., {Toon}, O.~B., \& {Colaprete}, A. 2008, \bibinfo{title}{{Modeling the environmental effects of moderate-sized impacts on Mars},} Journal of Geophysical Research (Planets), 113, E11007, \dodoi{10.1029/2008JE003147}

\bibitem[{T.~L. {Segura} {et~al.}(2002){Segura}, {Toon}, {Colaprete}, \& {Zahnle}}]{Segura2002}
{Segura}, T.~L., {Toon}, O.~B., {Colaprete}, A., \& {Zahnle}, K. 2002, \bibinfo{title}{{Environmental Effects of Large Impacts on Mars},} Science, 298, 1977, \dodoi{10.1126/science.1073586}

\bibitem[{A.~L. {Shields} {et~al.}(2016){Shields}, {Ballard}, \& {Johnson}}]{Shields2016}
{Shields}, A.~L., {Ballard}, S., \& {Johnson}, J.~A. 2016, \bibinfo{title}{{The habitability of planets orbiting M-dwarf stars},} \physrep, 663, 1, \dodoi{10.1016/j.physrep.2016.10.003}

\bibitem[{E.~M. {Shoemaker} {et~al.}(1969){Shoemaker}, {Batson}, {Holt}, {Morris}, {Rennilson}, \& {Whitaker}}]{Shoemaker1969}
{Shoemaker}, E.~M., {Batson}, R.~M., {Holt}, H.~E., {et~al.} 1969, \bibinfo{title}{{Observations of the lunar regolith and the earth from the television camera on Surveyor 7.},} \jgr, 74, 6081, \dodoi{10.1029/JB074i025p06081}

\bibitem[{E.~M. {Shoemaker} \& E.~C. {Morris}(1970){Shoemaker} \& {Morris}}]{Shoemaker1970}
{Shoemaker}, E.~M., \& {Morris}, E.~C. 1970, \bibinfo{title}{{Physical characteristics of the lunar regolith determined from Surveyor television observations.},} Radio Science, 5, 129, \dodoi{10.1029/RS005i002p00129}

\bibitem[{V. {Shuvalov}(2009){Shuvalov}}]{Shuvalov2009}
{Shuvalov}, V. 2009, \bibinfo{title}{{Atmospheric erosion induced by oblique impacts},} \maps, 44, 1095, \dodoi{10.1111/j.1945-5100.2009.tb01209.x}

\bibitem[{N.~H. {Sleep} \& K. {Zahnle}(1998){Sleep} \& {Zahnle}}]{SleepZahnle1998}
{Sleep}, N.~H., \& {Zahnle}, K. 1998, \bibinfo{title}{{Refugia from asteroid impacts on early Mars and the early Earth},} \jgr, 103, 28529, \dodoi{10.1029/98JE01809}

\bibitem[{A. {Soto} {et~al.}(2015){Soto}, {Mischna}, {Schneider}, {Lee}, \& {Richardson}}]{Soto2015}
{Soto}, A., {Mischna}, M., {Schneider}, T., {Lee}, C., \& {Richardson}, M. 2015, \bibinfo{title}{{Martian atmospheric collapse: Idealized GCM studies},} \icarus, 250, 553, \dodoi{10.1016/j.icarus.2014.11.028}

\bibitem[{R.~G. {Strom} {et~al.}(2008){Strom}, {Chapman}, {Merline}, {Solomon}, \& {Head}}]{Strom2008}
{Strom}, R.~G., {Chapman}, C.~R., {Merline}, W.~J., {Solomon}, S.~C., \& {Head}, J.~W. 2008, \bibinfo{title}{{Mercury Cratering Record Viewed from MESSENGER{\textquoteright}s First Flyby},} Science, 321, 79, \dodoi{10.1126/science.1159317}

\bibitem[{J.~S. {Stuart} \& R.~P. {Binzel}(2004){Stuart} \& {Binzel}}]{Binzel2004}
{Stuart}, J.~S., \& {Binzel}, R.~P. 2004, \bibinfo{title}{{Bias-corrected population, size distribution, and impact hazard for the near-Earth objects},} \icarus, 170, 295, \dodoi{10.1016/j.icarus.2004.03.018}

\bibitem[{F. {Tian}(2009){Tian}}]{Tian2009}
{Tian}, F. 2009, \bibinfo{title}{{Thermal Escape from Super Earth Atmospheres in the Habitable Zones of M Stars},} \apj, 703, 905, \dodoi{10.1088/0004-637X/703/1/905}

\bibitem[{O.~B. {Toon} {et~al.}(2010){Toon}, {Segura}, \& {Zahnle}}]{Toon2010}
{Toon}, O.~B., {Segura}, T., \& {Zahnle}, K. 2010, \bibinfo{title}{{The Formation of Martian River Valleys by Impacts},} Annual Review of Earth and Planetary Sciences, 38, 303, \dodoi{10.1146/annurev-earth-040809-152354}

\bibitem[{C.~C.~C. Tsang {et~al.}(2016)Tsang, Spencer, Lellouch, Lopez-Valverde, \& Richter}]{Tsang2016}
Tsang, C. C.~C., Spencer, J.~R., Lellouch, E., Lopez-Valverde, M.~A., \& Richter, M.~J. 2016, \bibinfo{title}{The collapse of Io's primary atmosphere in Jupiter eclipse,} Journal of Geophysical Research: Planets, 121, 1400, \dodoi{https://doi.org/10.1002/2016JE005025}

\bibitem[{M. {Turbet} {et~al.}(2020){Turbet}, {Gillmann}, {Forget}, {Baudin}, {Palumbo}, {Head}, \& {Karatekin}}]{Turbet2020}
{Turbet}, M., {Gillmann}, C., {Forget}, F., {et~al.} 2020, \bibinfo{title}{{The environmental effects of very large bolide impacts on early Mars explored with a hierarchy of numerical models},} \icarus, 335, 113419, \dodoi{10.1016/j.icarus.2019.113419}

\bibitem[{J.~C.~G. {Walker} {et~al.}(1981){Walker}, {Hays}, \& {Kasting}}]{Walker1981}
{Walker}, J.~C.~G., {Hays}, P.~B., \& {Kasting}, J.~F. 1981, \bibinfo{title}{{A negative feedback mechanism for the long-term stabilization of the earth's surface temperature},} \jgr, 86, 9776, \dodoi{10.1029/JC086iC10p09776}

\bibitem[{K. Wallmann \& G. Aloisi(2012)Wallmann \& Aloisi}]{WallmannAloisi2012}
Wallmann, K., \& Aloisi, G. 2012, \bibinfo{title}{The Global Carbon Cycle: Geological Processes,} in Fundamentals of Geobiology, ed. A.~H. Knoll, D.~E. Canfield, \& K.~O. Konhauser (Chichester, UK: Wiley-Blackwell), 20--38, \dodoi{10.1002/9781118280874.ch3}

\bibitem[{A.~J. {Watson} {et~al.}(1981){Watson}, {Donahue}, \& {Walker}}]{Watson1981}
{Watson}, A.~J., {Donahue}, T.~M., \& {Walker}, J.~C.~G. 1981, \bibinfo{title}{{The dynamics of a rapidly escaping atmosphere: Applications to the evolution of Earth and Venus},} \icarus, 48, 150, \dodoi{10.1016/0019-1035(81)90101-9}

\bibitem[{M. {Weiner Mansfield} {et~al.}(2024){Weiner Mansfield}, {Xue}, {Zhang}, {Mahajan}, {Ih}, {Koll}, {Bean}, {Park Coy}, {Eastman}, {Kempton}, {Kite}, \& {Lunine}}]{Mansfield2024}
{Weiner Mansfield}, M., {Xue}, Q., {Zhang}, M., {et~al.} 2024, \bibinfo{title}{{No Thick Atmosphere on the Terrestrial Exoplanet Gl 486b},} arXiv e-prints, arXiv:2408.15123, \dodoi{10.48550/arXiv.2408.15123}

\bibitem[{J.~G. {Winters} {et~al.}(2019){Winters}, {Medina}, {Irwin}, {Charbonneau}, {Astudillo-Defru}, {Horch}, {Eastman}, {Vrijmoet}, {Henry}, {Diamond-Lowe}, {Winston}, {Barclay}, {Bonfils}, {Ricker}, {Vanderspek}, {Latham}, {Seager}, {Winn}, {Jenkins}, {Udry}, {Twicken}, {Teske}, {Tenenbaum}, {Pepe}, {Murgas}, {Muirhead}, {Mink}, {Lovis}, {Levine}, {L{\'e}pine}, {Jao}, {Henze}, {Fur{\'e}sz}, {Forveille}, {Figueira}, {Esquerdo}, {Dressing}, {D{\'\i}az}, {Delfosse}, {Burke}, {Bouchy}, {Berlind}, \& {Almenara}}]{Winters2019}
{Winters}, J.~G., {Medina}, A.~A., {Irwin}, J.~M., {et~al.} 2019, \bibinfo{title}{{Three Red Suns in the Sky: A Transiting, Terrestrial Planet in a Triple M-dwarf System at 6.9 pc},} \aj, 158, 152, \dodoi{10.3847/1538-3881/ab364d}

\bibitem[{J.~G. {Winters} {et~al.}(2022){Winters}, {Cloutier}, {Medina}, {Irwin}, {Charbonneau}, {Astudillo-Defru}, {Bonfils}, {Howard}, {Isaacson}, {Bean}, {Seifahrt}, {Teske}, {Eastman}, {Twicken}, {Collins}, {Jensen}, {Quinn}, {Payne}, {Kristiansen}, {Spencer}, {Vanderburg}, {Zechmeister}, {Weiss}, {Wang}, {Wang}, {Udry}, {Terentev}, {St{\"u}rmer}, {Stef{\'a}nsson}, {Shporer}, {Shectman}, {Sefako}, {Schwengeler}, {Schwarz}, {Scarsdale}, {Rubenzahl}, {Roy}, {Rosenthal}, {Robertson}, {Petigura}, {Pepe}, {Omohundro}, {Murphy}, {Murgas}, {Mo{\v{c}}nik}, {Montet}, {Mennickent}, {Mayo}, {Massey}, {Lubin}, {Lovis}, {Lewin}, {Kasper}, {Kane}, {Jenkins}, {Huber}, {Horne}, {Hill}, {Gorrini}, {Giacalone}, {Fulton}, {Forveille}, {Figueira}, {Fetherolf}, {Dressing}, {D{\'\i}az}, {Delfosse}, {Dalba}, {Dai}, {Cort{\'e}s}, {Crossfield}, {Crane}, {Conti}, {Collins}, {Chontos}, {Butler}, {Brown}, {Brady}, {Behmard}, {Beard}, {Batalha}, \& {Almenara}}]{Winters2022}
{Winters}, J.~G., {Cloutier}, R., {Medina}, A.~A., {et~al.} 2022, \bibinfo{title}{{A Second Planet Transiting LTT 1445A and a Determination of the Masses of Both Worlds},} \aj, 163, 168, \dodoi{10.3847/1538-3881/ac50a9}

\bibitem[{R. {Wordsworth}(2015){Wordsworth}}]{Wordsworth2015}
{Wordsworth}, R. 2015, \bibinfo{title}{{Atmospheric Heat Redistribution and Collapse on Tidally Locked Rocky Planets},} \apj, 806, 180, \dodoi{10.1088/0004-637X/806/2/180}

\bibitem[{R. {Wordsworth} {et~al.}(2010){Wordsworth}, {Forget}, \& {Eymet}}]{Wordsworth2010}
{Wordsworth}, R., {Forget}, F., \& {Eymet}, V. 2010, \bibinfo{title}{{Infrared collision-induced and far-line absorption in dense CO $_{2}$ atmospheres},} \icarus, 210, 992, \dodoi{10.1016/j.icarus.2010.06.010}

\bibitem[{R. {Wordsworth} \& L. {Kreidberg}(2022){Wordsworth} \& {Kreidberg}}]{WordsworthKreidberg2022}
{Wordsworth}, R., \& {Kreidberg}, L. 2022, \bibinfo{title}{{Atmospheres of Rocky Exoplanets},} \araa, 60, 159, \dodoi{10.1146/annurev-astro-052920-125632}

\bibitem[{R.~D. {Wordsworth} {et~al.}(2018){Wordsworth}, {Schaefer}, \& {Fischer}}]{Wordsworth2018}
{Wordsworth}, R.~D., {Schaefer}, L.~K., \& {Fischer}, R.~A. 2018, \bibinfo{title}{{Redox Evolution via Gravitational Differentiation on Low-mass Planets: Implications for Abiotic Oxygen, Water Loss, and Habitability},} \aj, 155, 195, \dodoi{10.3847/1538-3881/aab608}

\bibitem[{Q. {Xue} {et~al.}(2024){Xue}, {Bean}, {Zhang}, {Mahajan}, {Ih}, {Eastman}, {Lunine}, {Mansfield}, {Coy}, {Kempton}, {Koll}, \& {Kite}}]{Xue2024}
{Xue}, Q., {Bean}, J.~L., {Zhang}, M., {et~al.} 2024, \bibinfo{title}{{JWST Thermal Emission of the Terrestrial Exoplanet GJ 1132b},} \apjl, 973, L8, \dodoi{10.3847/2041-8213/ad72e9}

\bibitem[{Q. {Xue} {et~al.}(2025){Xue}, {Zhang}, {Coy}, {Brady}, {Ji}, {Bean}, {Radica}, {Seifahrt}, {Sturmer}, {Luque}, {Basant}, {Brown}, {Das}, {Kasper}, {Piaulet-Ghorayeb}, {Kempton}, \& {Kite}}]{Xue2025}
{Xue}, Q., {Zhang}, M., {Coy}, B.~P., {et~al.} 2025, \bibinfo{title}{{The JWST Rocky Worlds DDT Program reveals GJ 3929b to likely be a bare rock},} arXiv e-prints, arXiv:2508.12516, \dodoi{10.48550/arXiv.2508.12516}

\bibitem[{K.~J. {Zahnle} \& D.~C. {Catling}(2017){Zahnle} \& {Catling}}]{ZahnleCatling2017}
{Zahnle}, K.~J., \& {Catling}, D.~C. 2017, \bibinfo{title}{{The Cosmic Shoreline: The Evidence that Escape Determines which Planets Have Atmospheres, and what this May Mean for Proxima Centauri B},} \apj, 843, 122, \dodoi{10.3847/1538-4357/aa7846}

\bibitem[{M. {Zhang} {et~al.}(2024){Zhang}, {Hu}, {Inglis}, {Dai}, {Bean}, {Knutson}, {Lam}, {Goffo}, \& {Gandolfi}}]{Zhang2024}
{Zhang}, M., {Hu}, R., {Inglis}, J., {et~al.} 2024, \bibinfo{title}{{GJ 367b Is a Dark, Hot, Airless Sub-Earth},} \apjl, 961, L44, \dodoi{10.3847/2041-8213/ad1a07}

\bibitem[{S. {Zieba} {et~al.}(2023){Zieba}, {Kreidberg}, {Ducrot}, {Gillon}, {Morley}, {Schaefer}, {Tamburo}, {Koll}, {Lyu}, {Acu{\~n}a}, {Agol}, {Iyer}, {Hu}, {Lincowski}, {Meadows}, {Selsis}, {Bolmont}, {Mandell}, \& {Suissa}}]{Zieba2023}
{Zieba}, S., {Kreidberg}, L., {Ducrot}, E., {et~al.} 2023, \bibinfo{title}{{No thick carbon dioxide atmosphere on the rocky exoplanet TRAPPIST-1 c},} \nat, 620, 746, \dodoi{10.1038/s41586-023-06232-z}

\end{thebibliography}



\end{document}